# Price Discovery and the Accuracy of Consolidated Data Feeds in the U.S. Equity Markets*

Brian F. Tivnan[1,2,3], David Slater[1] ,James R. Thompson[1] , Tobin A. Bergen-Hill[1,] Carl D. Burke[1] , Shaun M. Brady[4], Matthew T. K. Koehler[1], Matthew T. McMahon[1] , Brendan F. Tivnan[2] and Jason Veneman[1,3]

**Abstract:** Both the scientific community and the popular press have paid much attention to the speed of the Securities Information Processor – the data feed consolidating all trades and quotes across the US stock market. Rather than the speed of the Securities Information Processor, or SIP, we focus here on its accuracy. Relying on Trade and Quote data, we provide various measures of SIP latency relative to high-speed data feeds between exchanges, known as direct feeds. We use first differences to highlight not only the divergence between the direct feeds and the SIP, but also the fundamental inaccuracy of the SIP. We find that as many as 60% or more of trades are reported out of sequence for stocks with high trade volume, therefore skewing simple measures such as returns. While not yet definitive, this analysis supports our preliminary conclusion that the underlying infrastructure of the SIP is currently unable to keep pace with the trading activity in today's stock market.

## 1. INTRODUCTION

The scientific community and the popular press have paid much attention to the speed of the Securities Information Processor (SIP) – the data feed consolidating all trade and quote messages across the US stock market - relative to other data services [1]. Here, we address its importance.

Elsewhere, we have focused on the importance of the SIP to a key measure of market quality – efficient, price discovery [2]. Here, we focus on extending that analysis to assess the accuracy of the SIP. We use Trade and Quote data to provide various measures of SIP latency relative to direct feeds – the high-speed, data feeds between exchanges. Using first differences, we highlight the fundamental inaccuracy of the SIP. We find that 60% or more of trades are reported out of sequence for stocks with high volume. This disordering of trades skews simple measures such as returns. While preliminary, this analysis supports our conclusion that the underlying infrastructure of the SIP is currently unable to keep pace with the trading activity in today's U.S. equity markets.

In the sections which follow, we provide an overview of the National Market System and the SIP while adding clarity to the debates surrounding the SIP. We summarize the relevant literature of previous attempts to clarify the role of the SIP within the National Market System as well as describe our subsequent contribution to this literature. We describe our methods and data in greater detail, followed by a presentation of our findings. We then conclude the paper with a brief discussion of the implications of our findings.

*1.1. Overview of the National Market System*

The National Market System (colloquially known as the "stock market") includes all market centers where investors can buy and sell shares of publicly traded companies. To facilitate the efficient exchange of capital and shares in the National Market System (NMS), each market center is required to publish both the *Best Bid* (i.e., the highest price at which an investor is willing to pay for a single share of a given stock) and the *Best Offer* (i.e., the lowest price at which an investor is willing to sell a single share of a given stock) as well as how many shares the investor is willing to sell at that price. Across the entirety of the NMS, the highest bid and the lowest offer comprise what is known as the *National Best Bid and Offer* (NBBO) and the difference between the Best Bid and Best Offer is known as the spread.

* This manuscript will appear in a forthcoming issue of the *Journal of Risk and Financial Management*.
[1] The MITRE Corporation
[2] Complex Systems Center, University of Vermont
[3] Corresponding authors: btivnan@mitre.org and veneman@mitre.org
[4] Previous member of The MITRE Corporation, now director of Center for Model-Based Regulation

The NBBO reflects a distillation of the order flow across all the stock exchanges comprising the NMS. Figure 1 provides a graphical depiction of the three major datacenters of the NMS, all of which are located in northern New Jersey. Table A.1 also depicts the distribution of the families of exchanges across these datacenters. As depicted in Figure 1, the communications infrastructure connects the datacenters by dedicated, high-speed networks known as *Direct Feeds* depicted in red and black, as well as by the *Security Information Processor* depicted in pink. The Direct Feeds provide high-speed communications channels (e.g., faster than 50% of the speed of light [3]) where all orders flow between the exchanges. The Securities Information Processor, or SIP, consolidates and distills all the order flow from the Direct Feeds to determine and disseminate the NBBO as well as report all completed trades. We developed and posted an animation to provide a more illustrative depiction of these interactive dynamics between the direct feeds and the SIP.[5]

We conclude this overview of the National Market System with a description of the two Trade Reporting Facilities; namely, the NYSE Trade Reporting Facility (NTRF) and the NASDAQ Trade Reporting Facility (QTRF). Both Trade Reporting Facilities consolidate the trade reports from all Alternative Trading Systems (colloquially known as "Dark Pools"). The NTRF consolidates Dark Pool trade reports of NYSE-listed stocks whereas the QTRF consolidates Dark Pool trade reports of NASDAQ-listed stocks.

Unlike an exchange which must display best, local quotes, a Dark Pool is a market center where trading occurs without displaying any local quotes. Prevailing regulations still require that price discovery at Dark Pools are driven by the NBBO - same as at each exchange. While not depicted in Figure 1, we include Dark Pools here in our analysis for completeness, since trade reports from Dark Pools contribute to the overall, message traffic consolidated and disseminated by the SIP.

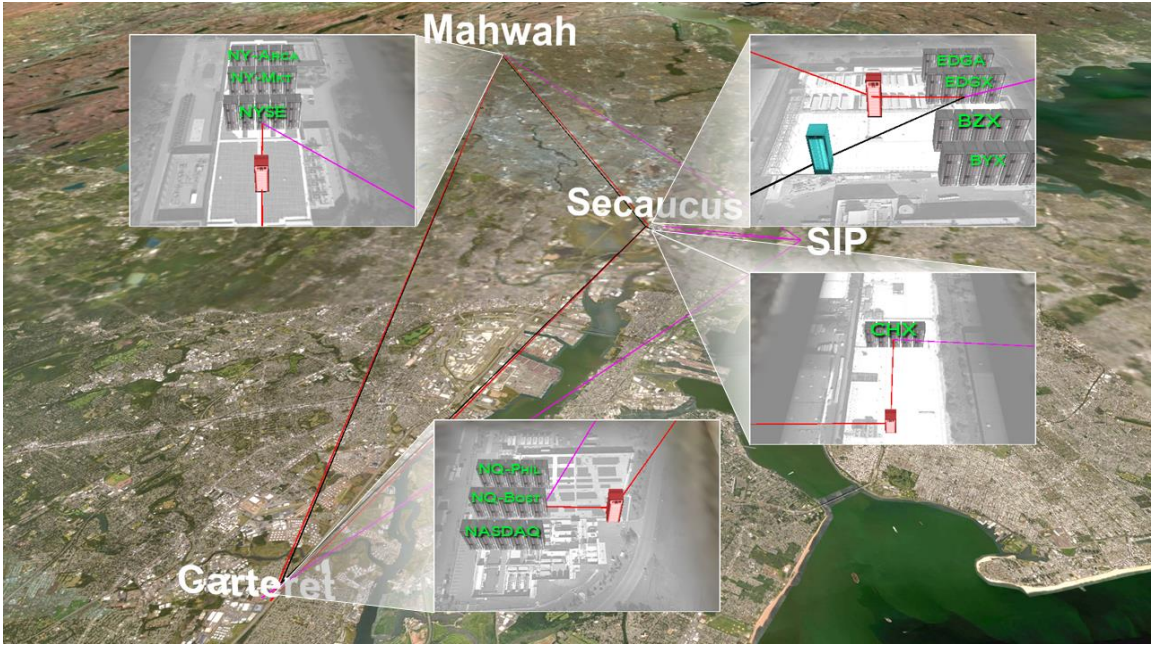

**Figure 1. Graphical depiction of the three major market centers in the National Market System.**

---

[5] The MITRE Corporation. "An Example of High Frequency Trader (HFT) Latency Arbitrage." Date of information retrieval: September 30, 2018.

## 2. REVIEW OF RELEVANT LITERATURE

There is widespread agreement that Mandelbrot [4] was one of the first to characterize the dynamic behavior of prices in modern markets; later to be extended by Cont [5]. Previously, we [6, 7] identified some anomalous behavior in the dynamics of market prices which deviated from previous expectations [4,5]. At the time, however, the resolution of our data prevented us from making any definitive determinations about the causal mechanisms of these behaviors.

Initially in the press and later in his book, Scott Patterson was one of the first to draw attention to the speed discrepancy between the SIP and the direct feeds [8, 9], only later to be followed in 2014 by Michael Lewis' [10] popular book, *Flash Boys*. Neither of these authors fully explored the implications of the speed discrepancy, instead relying heavily upon the works of financial industry practitioners.

A few industry practitioners were the first to identify the speed discrepancy between that of the SIP and the direct feeds. Arnuk and Saluzzi [11] were one of the first to highlight these latency discrepancies. Bodek [12] detailed the sweeping changes instituted by RegNMS, changes which stimulated the technological arms race and the insatiable demand for greater speeds from the direct feeds. And Nanex Research [13] highlighted the spoils which can be extracted in specific instances by those market participants so fortunate as to subscribe to the direct feeds.

Scientific contributions to this debate have been preliminary, possibly due to the limited availability of authoritative data within the scientific community. While authoritative datasets exist, subscriptions to them are extraordinarily expensive and therefore largely available only to specialized, market participants along with regulatory agencies.

Ding et al [14] provide one of the first scientific analyses of direct feed and SIP latencies. More recently, Bartlett and McCrary [15] attempted to revise and generalize earlier findings [13] by quantifying the total profits from subscribing to the direct feeds.

One common theme prevails throughout these analyses: the implication that the SIP accurately depicts market dynamics, albeit subject to some latency. We address this limitation of previous studies by highlighting and quantifying the inaccuracies of the SIP.

## 3. METHODS

The implications of ever increasing market speeds can be a difficult topic to understand even for those immersed in market data daily. The following is an attempt to break down some empirical analyses on granular data from US stock markets that illustrate some of the implications from SIP latencies and the subsequent impact on the accuracy of the SIP. We rely on Trade and Quote data spanning the National Market System. More precisely, we purchased the NxCore dataset from Nanex Research for the following periods: calendar year 2014 and the second and third quarters of 2015.

Many of the subsequent examples stem from market activity on a representative day (i.e., August 11th, 2015), which we chose for three reasons.  First, August 11th, 2015 was a not a noteworthy day in the U.S. equity markets (i.e., the absence of any market-wide events).  Second, this representative day closely precedes a notable day in the U.S. equity markets (i.e., August 24th, 2015), which is known colloquially as "Manic Monday."  Third, there were no announced modifications to the SIP between August 11th and 24th.  Given these three reasons above, it is realistic to assume the same underlying, market infrastructure for both days.  Therefore, August 11th, 2015 serves as a useful proxy for a representative day in the U.S. equity markets.

We capture both breadth and depth in our analyses. We achieve breadth in two ways.  First, we analyzed the entire population of the 7,993 unique tickers (i.e., stocks and exchange-traded funds) that printed quotes and trade reports to the SIP on the representative day.  The reader will note that an exchange-traded fund (ETF) is a security similar to a mutual fund, but an ETF trades throughout the day in the same manner as a stock. Second, we also closely analyzed SPY (i.e., the SPDR S&P 500 ETF).  It is important to note that the SPY is often used as a proxy of the U.S. equity market.  Unlike the S&P 500 index which cannot be directly traded, the SPY represents one of the largest and most liquid, market-wide securities.  For the SPY, we analyzed price, spread and market crosses.  A special case of the spread occurs when the spread becomes negative.  Such instances of a negative spread for the NBBO depict a *crossed market*, or simply *cross* for short.  While a small spread indicates general agreement on price across all market participants, a large spread and crosses indicate little agreement on price.  Both large spreads and crosses are indicators of inefficient, price discovery.

We also achieve depth of our analysis in a comprehensive manner. Rather than limiting our analyses to any one stock or relying solely on the large-cap stocks that comprise the Dow 30 as in previous studies [15], we chose a convenience sample of 14 stocks. This heterogeneous set of stocks varies on market capitalization, listed exchange and trading volume. This sample ranges from Apple (AAPL), one of the largest and most actively traded stocks in the world to Acme United Corporation (ACU), which is a lightly traded, small-cap stock.

Our analyses include some depictions of common descriptive measures of market performance, namely price and trade volume, as well as the monetary value of that trading activity. We then de-construct this market activity by specific exchange and components of the SIP.

We continue our analyses with a focus on the latencies between the direct feeds and the SIP, beginning with simple counts and frequency distributions. We extend this analysis by comparing first differences to highlight the divergence between the direct feeds and the SIP. We conclude with an analysis of these measures for comparison with known market events.

## 4. RESULTS

As described above, we begin with empirical evidence confirming our focal date as a representative day in the U.S. equity markets. From there, we provide depictions of common descriptive measures of market performance. We then de-construct this market activity by specific exchange and components of the SIP. We then present our analysis of first differences to highlight the divergence between the direct feeds and the SIP. We conclude with an analysis of these measures for comparison with known, market events.

*4.1. Representative Day in the U.S. Equity Markets*

For simplicity and clarity in the presentation of our analyses, we chose a representative day in the U.S. equity markets (i.e., August 11th, 2015) for its distinction from a noteworthy day (i.e., August 24th, 2015) within only a two-week span.  Figure 2 and Figure 3 provide the empirical evidence to support this distinction.  Both figures plot the price, spread and market crosses for SPY.

Figure 2 depicts the market dynamics for SPY on August 24th, 2015.  Leading up to and through the opening of the market day, the widening spread and frequent crosses indicate a tremendous uncertainty of the price.  Even as spreads narrowed and price stabilized throughout the market day, frequent crosses persisted.  These are all indicators of inefficient, price discovery.

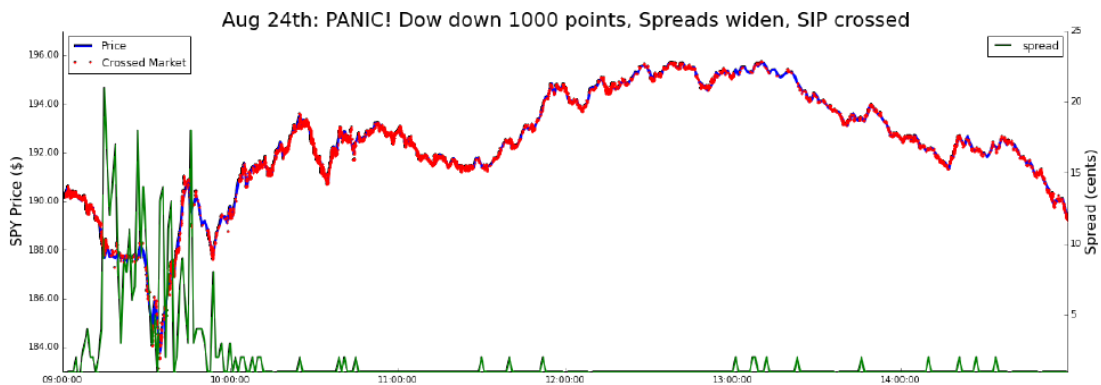

**Figure 2. Crosses and Spreads in the SPY on August 24th, 2015 – "Manic Monday."**

Figure 3 depicts the market dynamics for SPY on August 11th, 2015.  Unlike the dynamics of "Manic Monday," Figure 3 reflects a "typical market day."  More precisely, spreads remain small throughout the market day and crosses are intermittent yet fleeting while price is largely stable.

Moreover, Figure 3 provides no evidence that would lead one to conclude that the underlying, market infrastructure is overwhelmed in any meaningful way on this day.

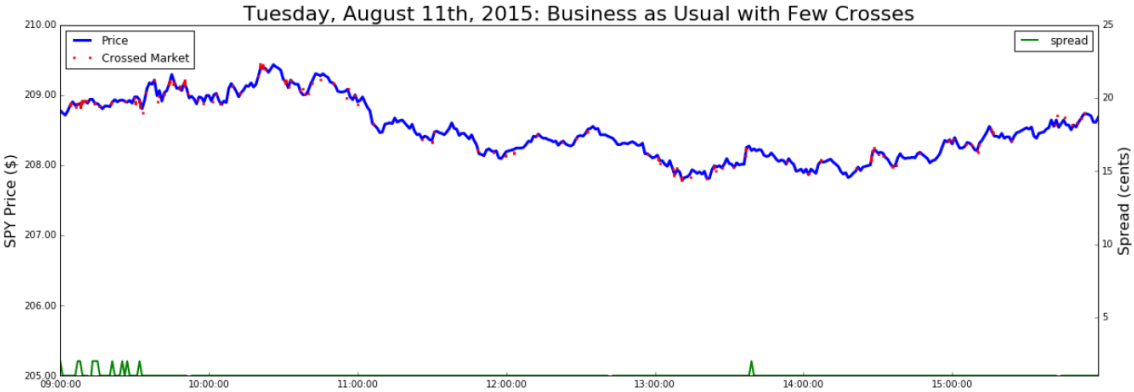

**Figure 3. Crosses and Spreads in the SPY on August 11th, 2015 – "typical market day."**

*4.2. Common, Descriptive Measures of Market Activity*

We naturally follow with a time-series depicting the price of a share of Apple over the course of a day. Figure 4 shows the price of Apple, including both pre-market (04:00 to 09:30 hours or 0 to 19,800 seconds in the figures below) and after-hours trading (16:00 to 20:00 hours or 43,200 to 57,600 seconds in the figures below). Figure 4 is typical of the only chart most people see when looking at the stock market. With data at the microsecond level, even in this typical view, one starts to see some anomalous behavior with the spikes at around 44,000 seconds. Those spikes are after-hours trades which, along with pre-market, are not usually shown on popular websites (e.g., Yahoo Finance). This begins to illustrate the complexity dynamics of price behaviors across an entire day, not just during the middle of the day.

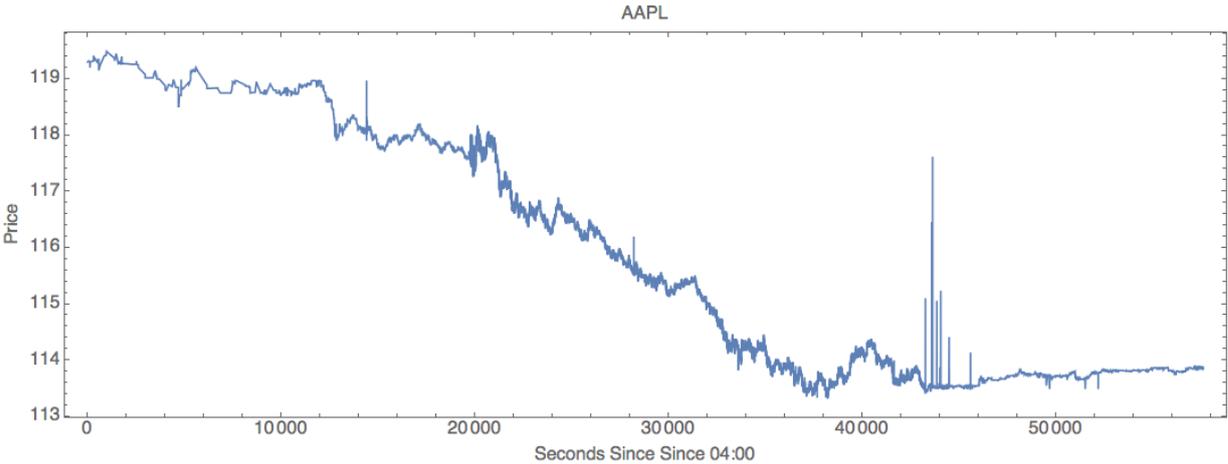

**Figure 4. Apple stock price on August 11th, 2015 at each second of the day from 04:00 until 20:00.**

In Figure 5, we illuminate the interesting dynamics associated with the number of trades per second for a single stock. Here, one will notice the clear difference in activity during pre-market, regular market (09:30 – 16:00), and after-hours trading (16:00 – 20:00).

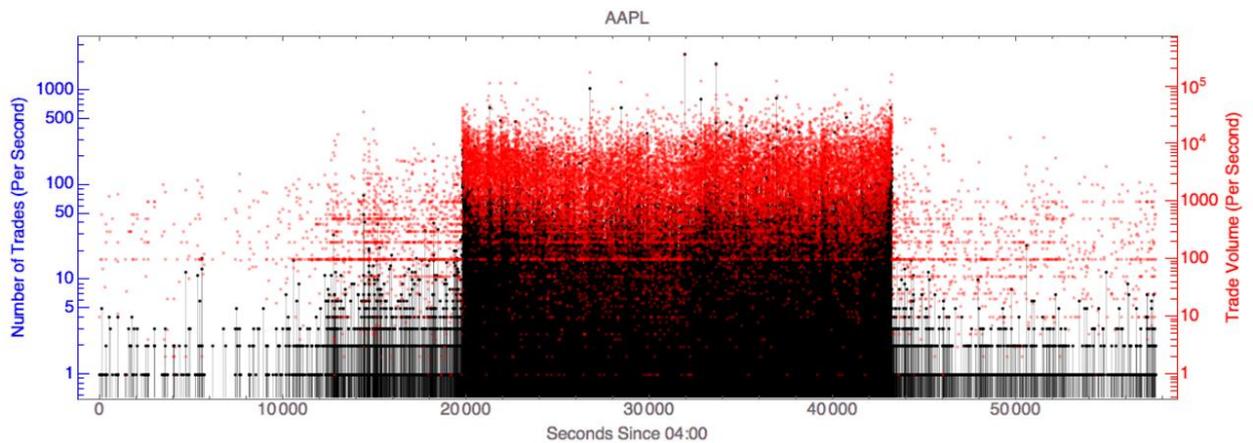

**Figure 5. Trades in Apple stock on August 11th, 2015 at each second of the day from 04:00 until 20:00.**

To get a sense of how much capital is circulating, a look into the number of dollars traded per second in Figure 6 reveals some astounding numbers. Here, one sees that at the highest spike $40 million in Apple stock is traded in one second.

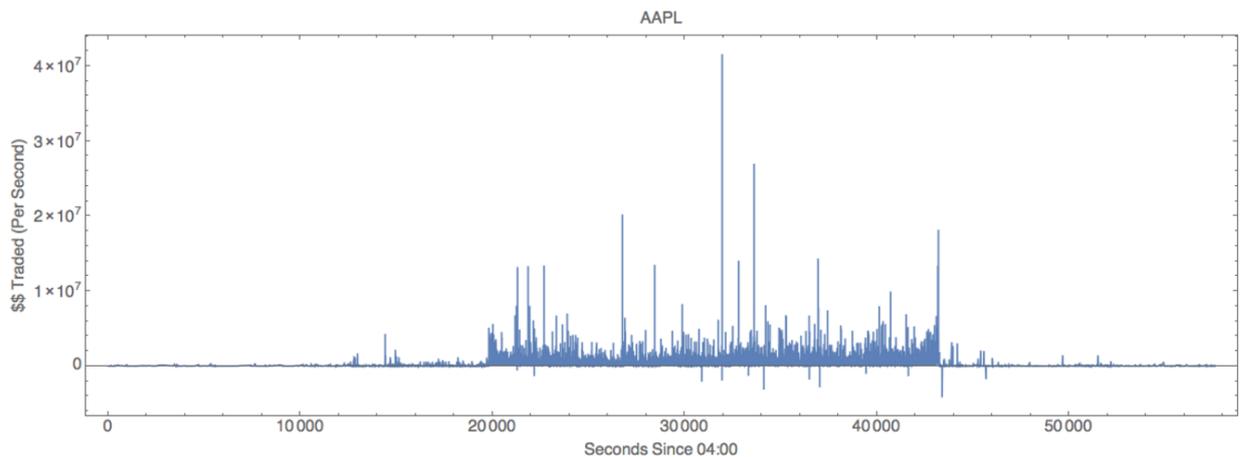

**Figure 6. Dollars traded in Apple on August 11th, 2015 at each second of the day from 04:00 until 20:00.**

To put things in perspective the cumulative traded capital in Figure 7 shows a steady climb to approximately $10 billion traded in a single day for Apple stock. At such a high rate, the $40 million spikes seen in Figure 6 above hardly register as outliers.

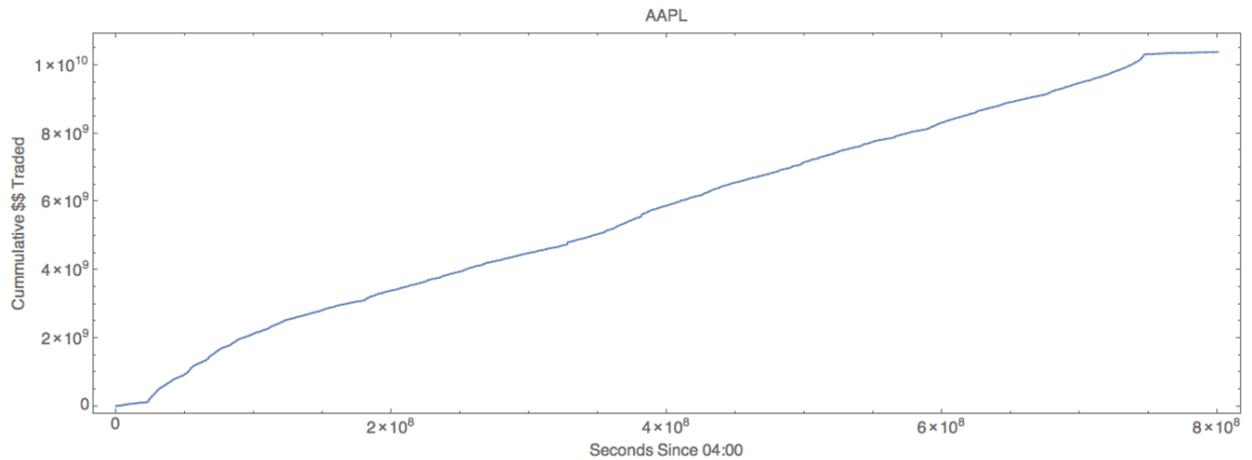

**Figure 7. Cumulative dollars traded in Apple stock on August 11th, 2015 summed each second from 04:00 until 20:00.**

*4.3. De-construct Market Activity by Exchange and Components of the SIP*

Figure 8 shows how assets are traded at a high volume at the multiple exchanges that comprise the National Market System. The opening and closing hours differ per exchange and are the reason that some lines start late or stop short. The start of regular trading at 09:30 (19,800 seconds since 04:00) is seen in Figure 8 as the sharp increase in volume at all markets at opening time.

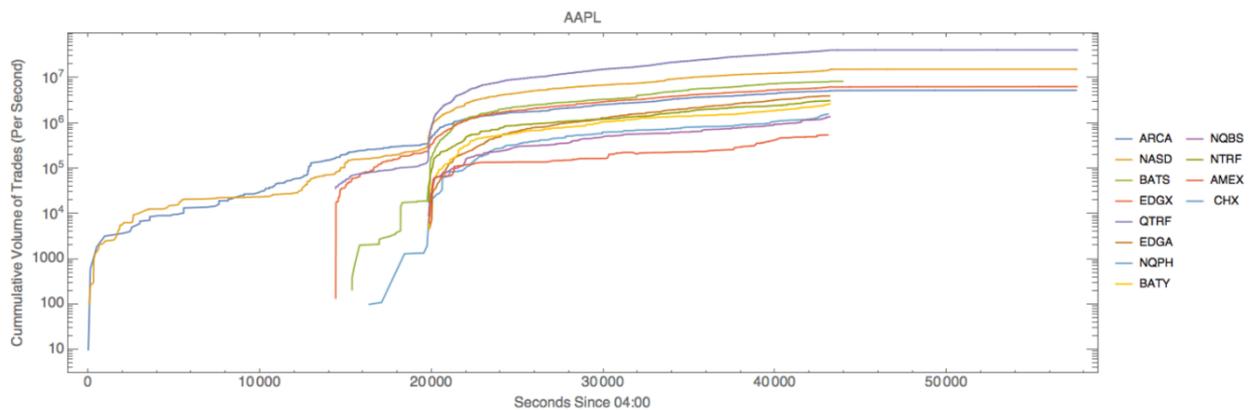

**Figure 8. Cumulative trade volume by exchange for Apple stock on August 11th, 2015 summed each second from 04:00 until 20:00.**

Here, we begin to tie the above financial measures with the concept of bandwidth capacity of the underlying communications networks, it may be more intuitive to think of trade volume and quote lots (a quote lot is typically 100 shares) in terms of messages. A message is defined as an atomic unit of communication that describes a number of shares or lots for a particular asset (e.g., 100 shares of Apple stock). Since there are far more quote messages than there are trade messages (i.e., roughly 10 quote messages for each trade report), the following figures of quote messages per day represent the upper bound of traffic on communication networks.

For simplicity, we have referred to the SIP as a single entity to this point in our analysis. In fact, the SIP is comprised of three, distinct components: SIP A, SIP B and SIP C. The SIPs link "the U.S. markets by processing and consolidating all protected bid/ask quotes and trades from every trading venue into a single, easily consumed data feed [16]." SIP A and B are operated by the Consolidated Tape Association and consolidate market activity for securities listed on New York Stock Exchange (NYSE) on SIP A and securities listed on NYSE ARCA, NYSE MKT, BATS and regional exchanges on SIP B [16]. SIP C is operated by NASDAQ for NASDAQ-listed securities [17].

In Figure 9, one can again clearly see the opening and closing of the regular market day. Figure 9 shows messages that are recorded at each of the three SIPs. During the market day, there are very few instances that less than 1,000 messages per second are printed to the SIPs, while there are times such as the market open and close when message traffic exceeds 100,000 messages per second.

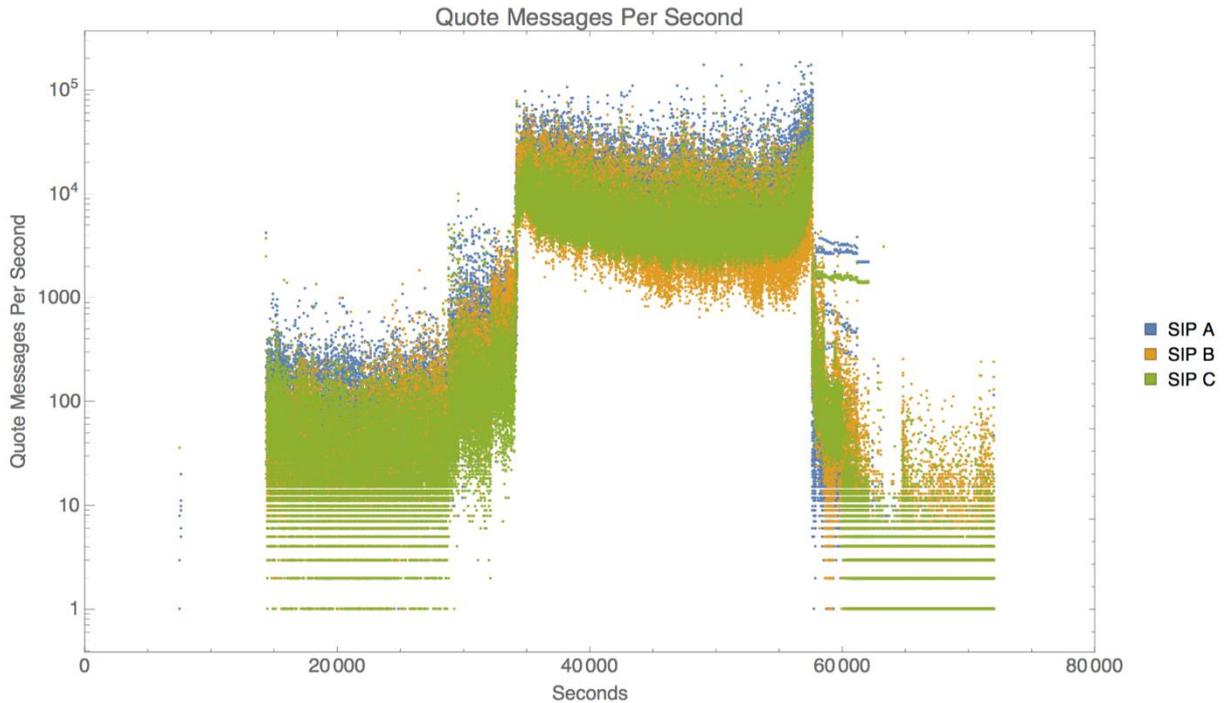

**Figure 9. Quote messages by SIP for Apple stock on August 11th, 2015 for each second from 04:00 until 20:00.**

*4.4. Latencies between the Direct Feeds and the SIP*

Recall that the SIP exists to unify the market in time via efficient discovery of equity prices. To get a picture of how long it takes for trades to go from being reported at an exchange to being aggregated and recorded by the SIP for all market participants to see, one can look at the SIP latency. Latency is defined here by the timestamp from the SIP minus the timestamp from when the reporting exchange sent the trade or quote information to the SIP. The latency in microseconds for trades at each of the three SIPs for all stocks is shown in Figures Figure 10, Figure 11 and Figure 12.

From these three figures, one can see that the median, standard deviation and range in latency are similar across each SIP. The median latencies across all SIPs and all exchanges are within 700μs of each other. From this, one might infer that trade reporting latencies are similar.

While similarities in latency median, standard deviation and range certainly do exist, there are at least two exceptions to these similarities. The first exception is the Chicago Stock Exchange (CHX). It was first reported elsewhere that CHX latencies are outliers from that of the other exchanges, as well as that CHX has limited share of the trading volume despite often contributing to efficient, price discovery [20]. Our findings here confirm CHX latencies as outliers. The second exception relates to the NTRF where Dark Pool trades are consolidated for NYSE-listed stocks.

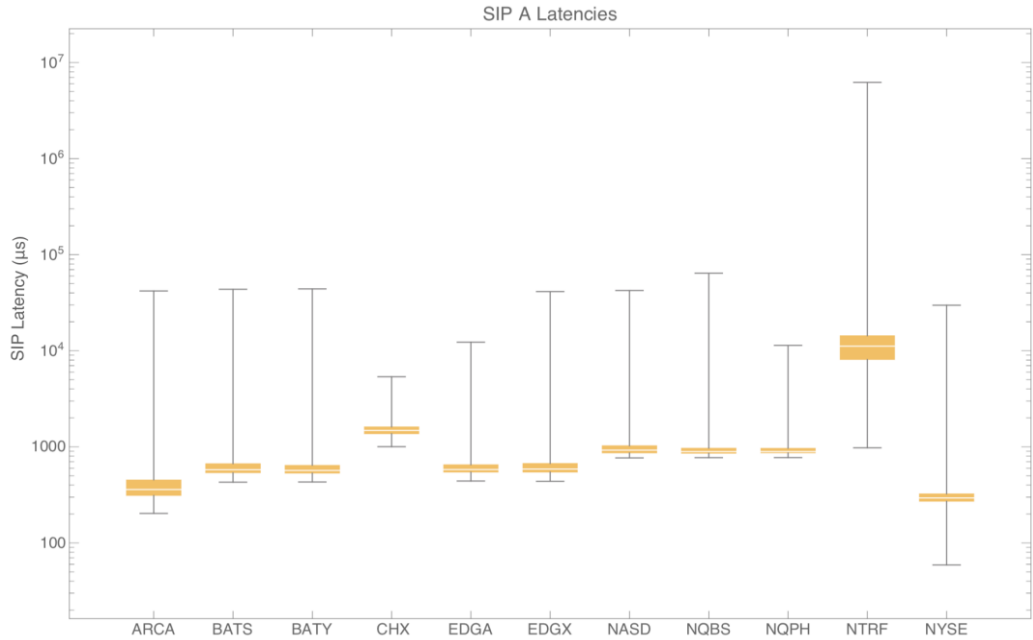

**Figure 10. Box plots of latencies for all trades of all stocks traded on August 11th, 2015 in microseconds reported by SIP A and separated by exchange.**

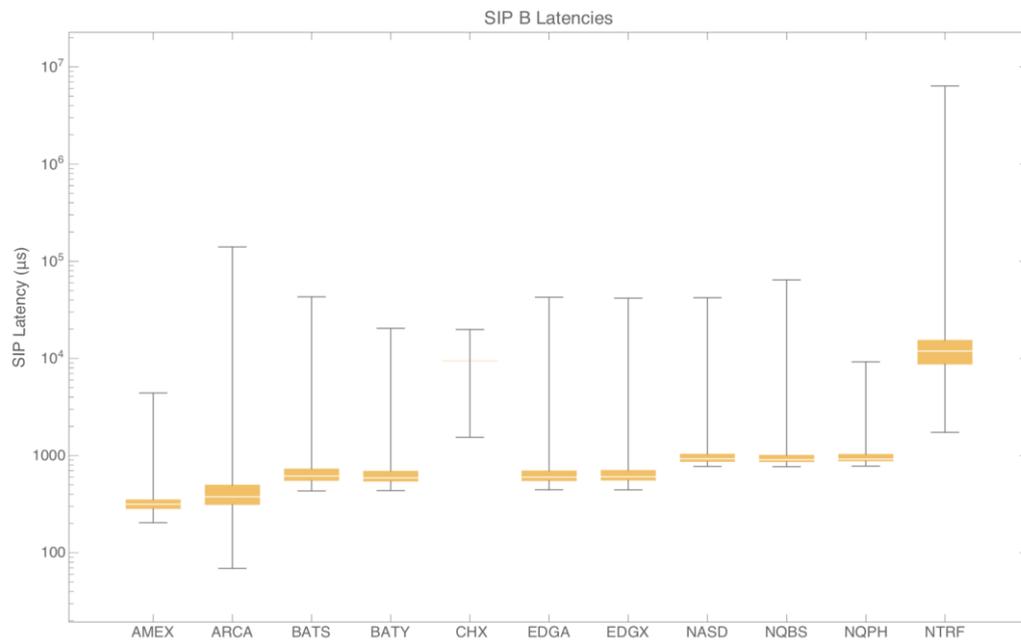

**Figure 11. Box plots of latencies for all trades of all stocks traded on August 11th, 2015 in microseconds reported by SIP B and separated by exchange.**

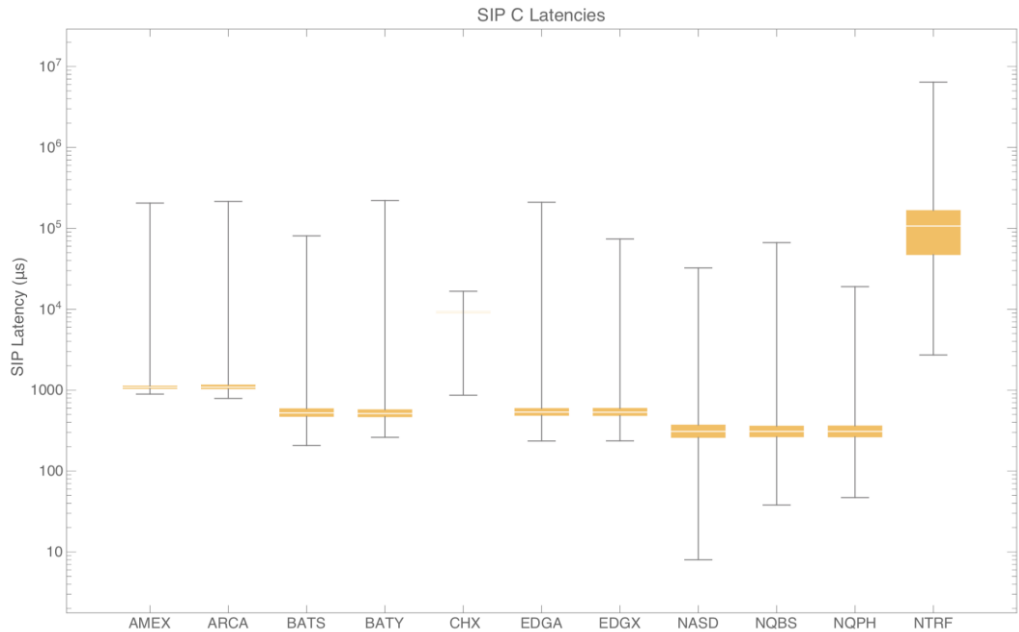

**Figure 12. Box plots of latencies for all trades of all stocks traded on August 11th, 2015 in microseconds reported by SIP C and separated by exchange.**

Looking at latency in more detail, Figure 13 shows the latency in trade reporting for a single stock (i.e., AAPL ). Figure 13 presents a far different picture from the SIP latencies and one can see that the latency differs significantly within and between exchanges. The outlier reporting exchange here is QTRF. This clearly shows that the exchanges matter when observing latency.

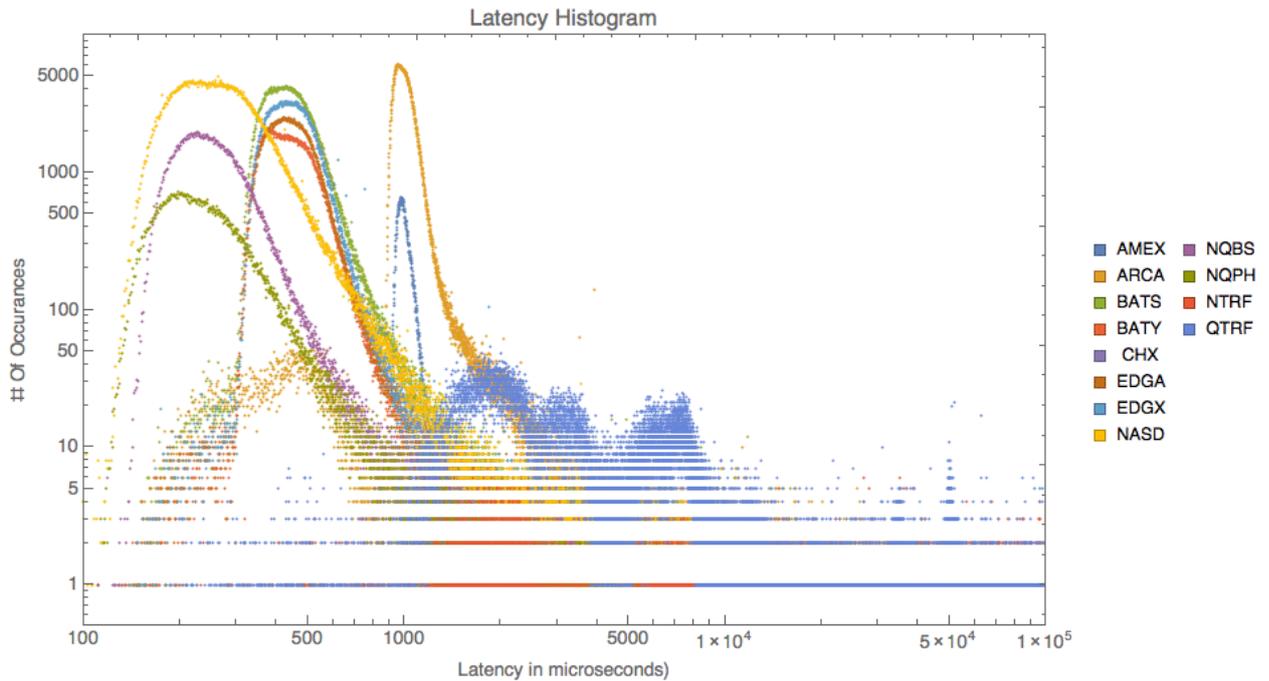

**Figure 13. Latency histogram for Apple stock traded on August 11th, 2015 by exchange for AAPL (truncated at $10^5$ microseconds).**

*4.5. First Differences Highlight Inaccuracy of the SIP*

Figure 14 shows the difference between the best bid and best ask, (i.e., the spread). We generated Figure 14 from the timestamps of the bid and ask quotes as printed to the SIP. This analysis shows that there are a non-trivial number of instances where the spread is negative. Indicative of inefficient price discovery, a negative spread means that one is willing to pay more than a counterparty is willing to sell at that specific time (e.g., buying a soda for $1.05 when the sticker price is only $0.99). This is neither desirable for the buyer nor is it allowed by current market rules [18], so the SIP might be disseminating inaccurate reports. To assess the accuracy of the SIP, one can look at the spread at each exchange in Figure 15 and see that at any exchange the spread never goes negative. However, instances of a negative spread do occur, particularly a negative spread for the NBBO. Such instances of a negative spread for the NBBO depict a *crossed market*, or simply *cross* for short. For completeness, a spread of zero, known as a *locked* market, depicts another market condition precluded by current market rules. All available orders at the best bid should have been executed against all available orders at the best offer.

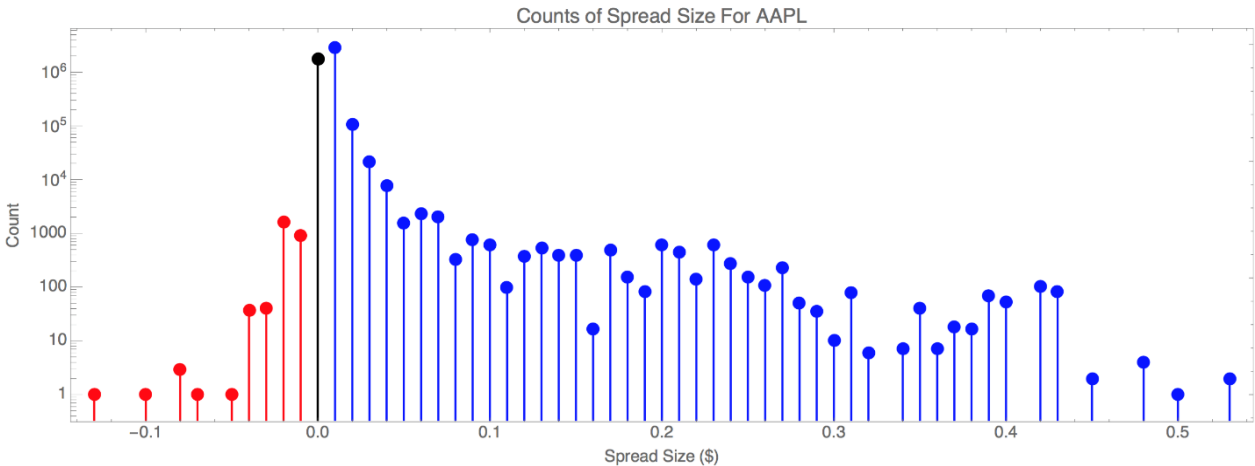

**Figure 14. Bid/ask spread in dollars vs. number of occurrences for Apple stock on August 11th, 2015.**

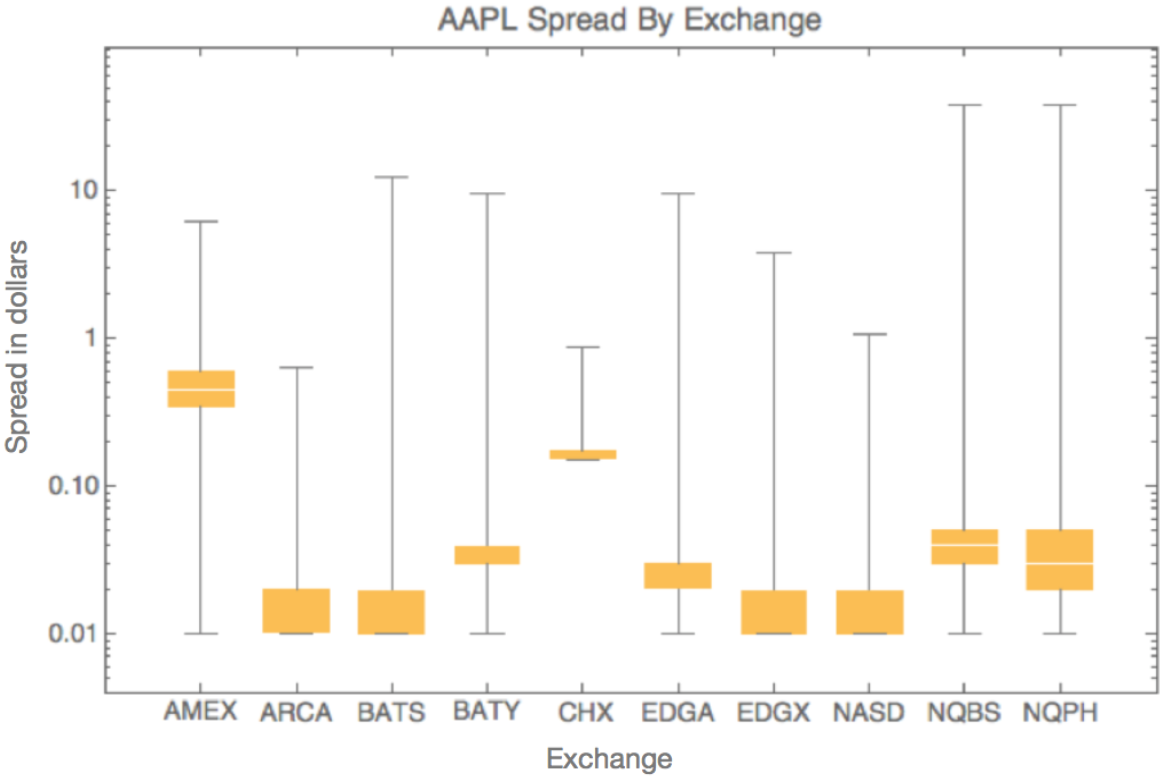

**Figure 15. Box plots of spread in dollars for Apple on August 11, 2015 by exchange.**

If the SIP appears wrong and the exchanges are never crossed, then this discrepancy may be due to the time it takes for messages to travel from an exchange to the SIP along communications networks. The presence of timing related anomalies indicates the importance of understanding the coupling between the financial markets and communications infrastructure and the implications of how these systems are designed.

These pricing anomalies are present across many stocks (e.g., the component stocks of the Dow 30), but for brevity and space considerations, we limit our treatment here to a convenience sample of 14 stocks listed in Table 1. Some patterns are indicated that give preliminary evidence of a relationship between message traffic and apparent crosses and locks (equal bid and asks which are also disallowed by market rules). If this relationship holds it gives further weight to the importance between finance and communication infrastructure coupling.  Figure 16 indicates that as the number of quote messages increases, there is an increasing number of apparent market crossings as seen at the SIP. In Figure 16, the price of the stock is given in the color chart to the right. Stocks that trade for less than one dollar (i.e. penny stocks, blue to blue-green in the figure) operate under different market rules and appear to follow a different relationship between number of messages and crosses. This preliminary result could indicate that market rules also shape the coupling between these infrastructures.

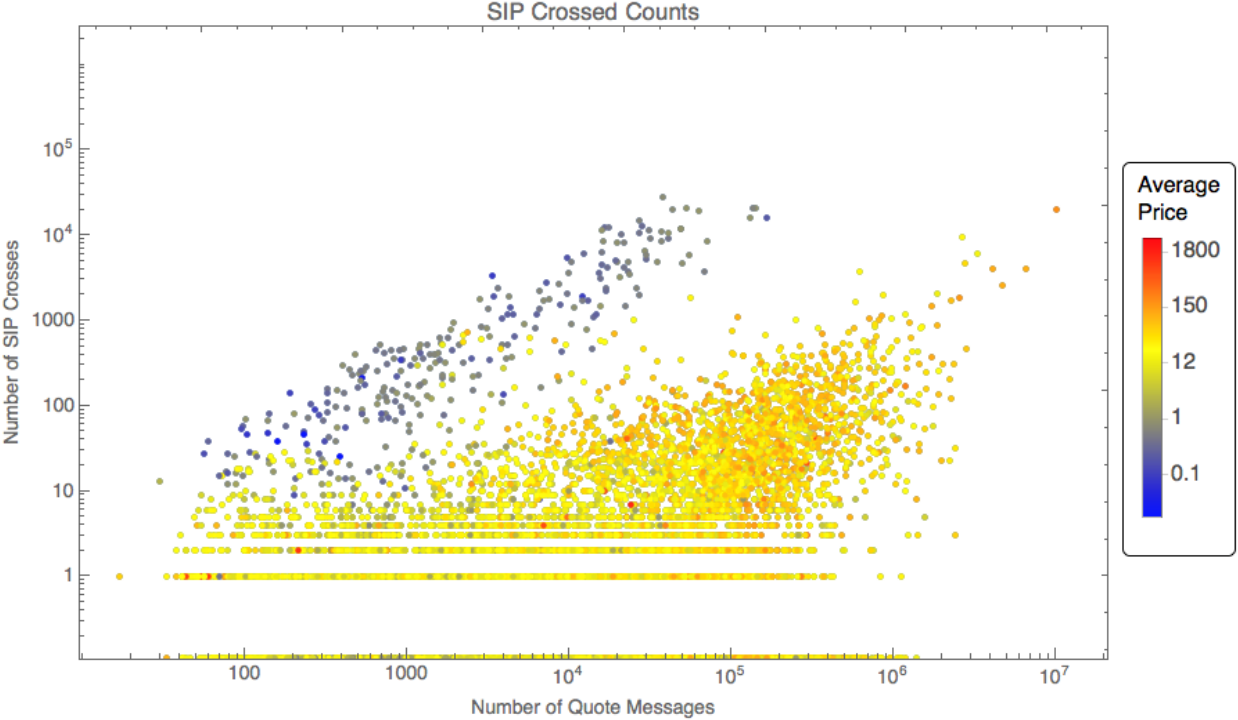

**Figure 16. Counts of SIP crosses for all stocks traded on August 11th, 2015.**

Figure 17 shows the number of times that a stock is apparently locked (i.e. a spread of 0) at the SIP. Figure 17 provides additional evidence that increasing message traffic correlates to more locked stocks.

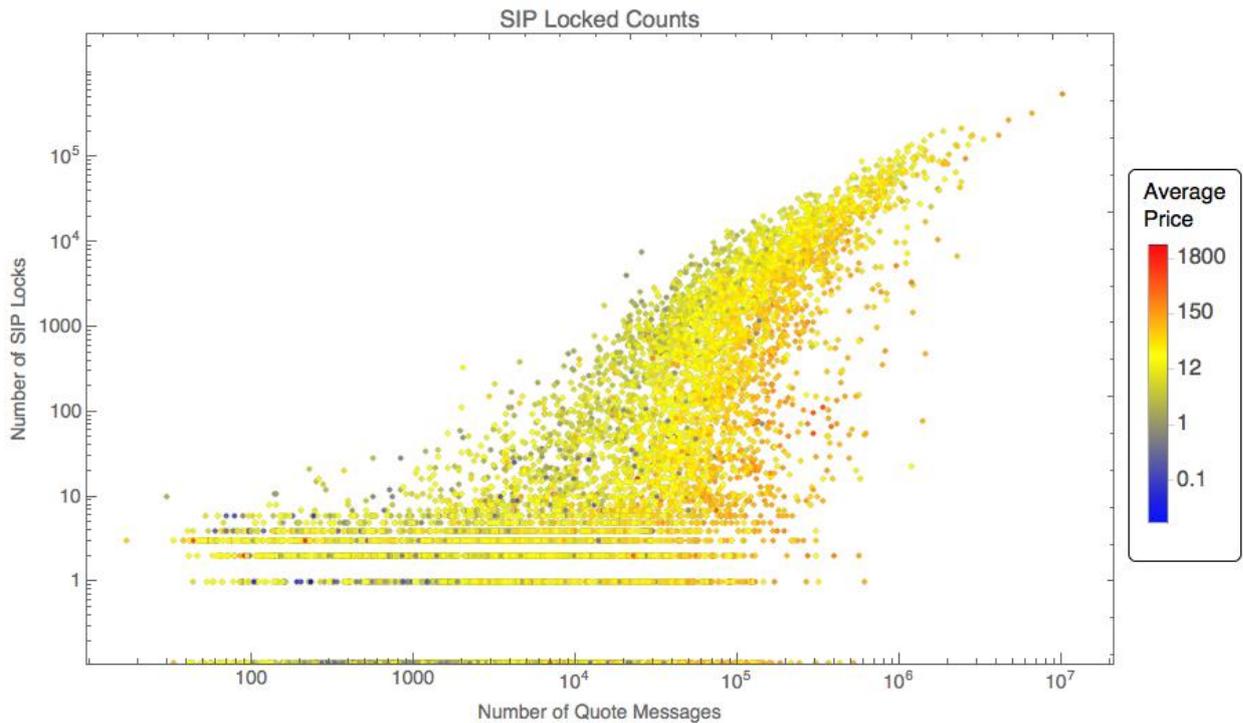

**Figure 17. Counts of SIP locks for all stocks traded on August 11th, 2015.**

The modern financial system relies on efficient and stable communication networks for dissemination of up-to-date information. To get a picture of what can happen in a short time window, one can look at the number of events (trade reports or quote messages) for a stock that occur between the time a trade or quote update is sent to the SIP and the time the SIP reports that update to SIP subscribers – here, we refer to this delay as the latency window. Figure 18 shows that it's very common to see from hundreds to thousands of events occur within this latency window for Apple stock.

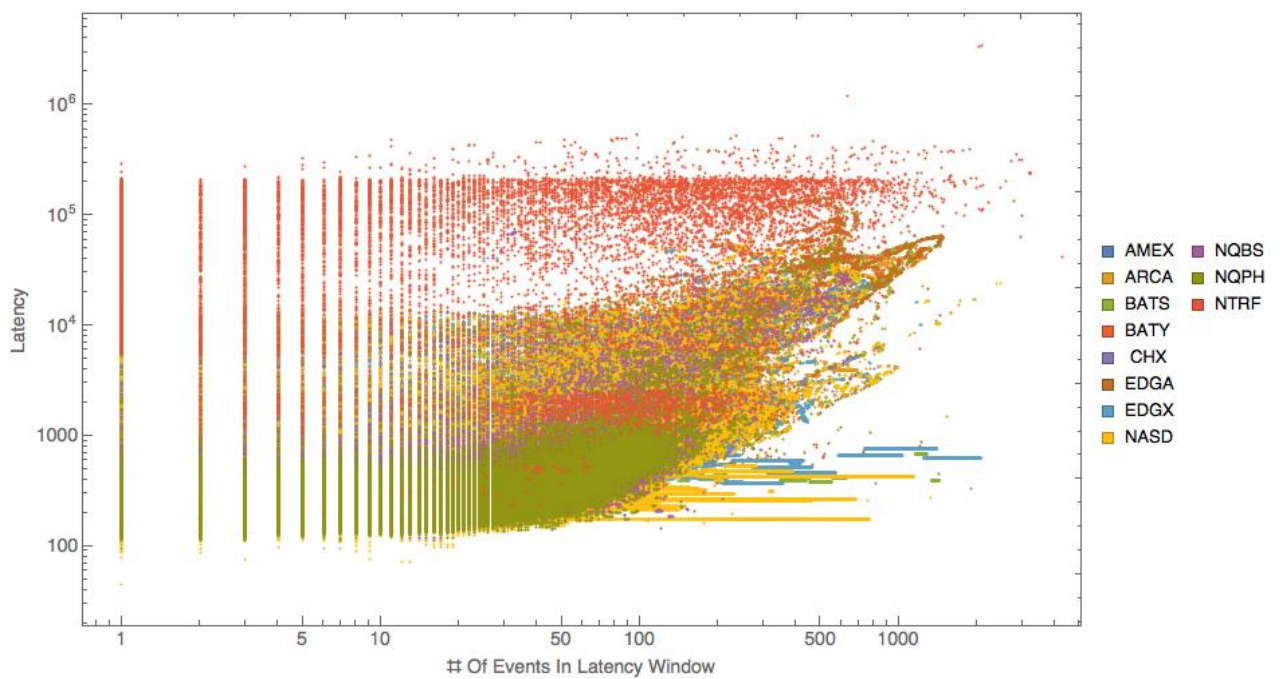

**Figure 18. Number of events in latency window for Apple stock on August 11th, 2015.**

*4.6. Crossings Between Order Books Present Arbitrage Opportunities*

The appearance of a relationship between the number of apparent market locks and crosses and message (trade report or quote update) volume prompts a deeper look into the number of events within latency windows. Figure 19 and Figure 20 are aligned vertically on the page to reflect time of day along the x axis. In Figure 19, we depict a lightly traded stock (i.e., Second Sight Medical Products, EYES) from our convenience sample to compare with a heavily traded stock (i.e., Apple, AAPL) in Figure 20. In Figure 19, the reader will notice that EYES has limited trading activity before and after the trading day, whereas Figure 20 depicts significant trading before and after hours in AAPL.

In Figure 19, we can see that there are some times where the SIP and exchange time do not line up (places where the red and white circles don't line up). For a highly traded stock like AAPL, we see from Figure 20 that there are far more occasions where the difference in time stamps is out of sequence.

A closer analysis of Figure 19 and Figure 20 yields additional insights. Because we sorted by the SIP timestamp using the Precision Time Protocol [19], all first differences should be positive. Therefore, all negative differences in the SIP timestamps, as depicted in Figures Figure 19 and Figure 20, indicate that the SIP may often misrepresent the accurate sequence of trades.

We see similar features in a number of stocks across a range of trading volume in Table 1 which shows the total number of trades for a particular asset and the percentage of those that were reported out of sequence by the SIP. A strong linear trend (slope = 0.66, R-squared = 0.99) exists between the number of trades and the number out of sequence. While additional analyses would be necessary to pinpoint drivers of that trend, such in-depth analyses could prove worthwhile as they might illuminate arbitrage opportunities. These arbitrage opportunities could arise from a crossed market. Recall that a crossed market indicates a negative spread (i.e., a displayed, best bid exceeds a displayed, best offer). If permitted, trading in a crossed market might present opportunities for true arbitrage. As an illustrative example, trading in a crossed market might enable Trader C to purchase an even lot of shares of Apple at $116.00 per share from Trader A and immediately sell the same even lot of Apple shares for $116.02 per share to Trader B. The reader will note that in the absence of Trader C in the market, Traders A and B might be natural counterparties at $116.01 per share of Apple, each experiencing price improvement which is why trading is halted in crossed markets. To exploit this potential arbitrage opportunity, Trader C must anticipate the crossed market and exploit a latency advantage via a direct-feed connection to the market. While market participants may have the ability to act on these arbitrage opportunities, they may exist for only a very small window of time (e.g., 500 microseconds or less) [2]. These fleeting opportunities for latency arbitrage clearly indicate the ramifications of the tight interdependence between financial markets and the communication networks which connect them.

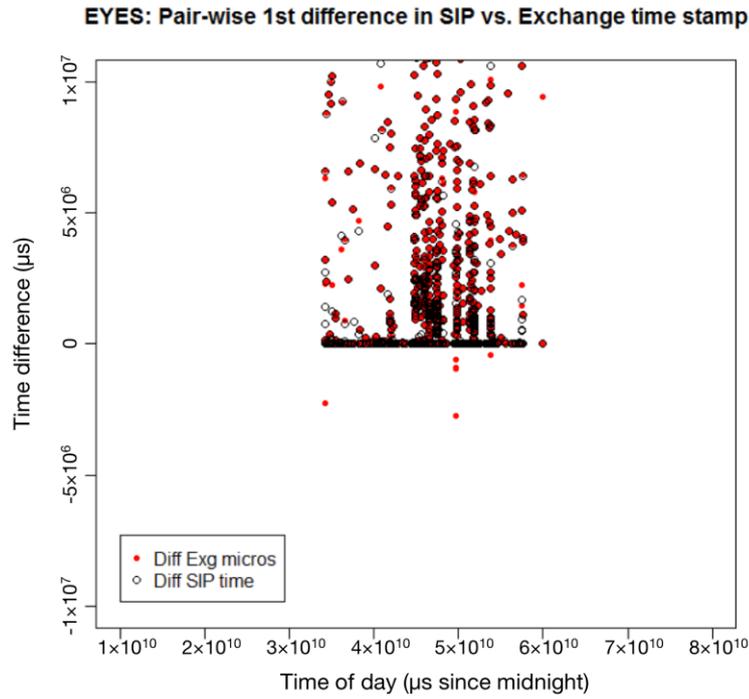

**Figure 19. Time difference in SIP vs. exchange time stamp for EYES stock on August 11th, 2015.**

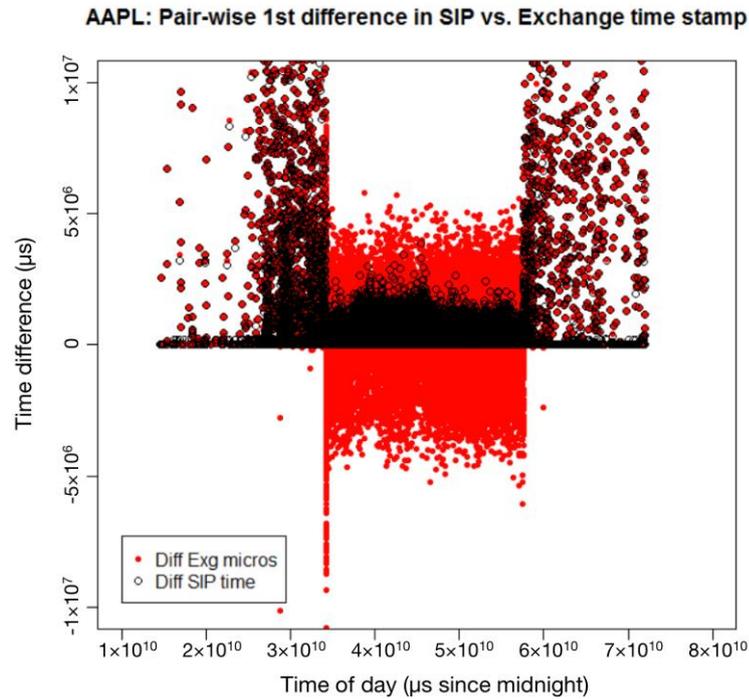

**Figure 20. Time difference in SIP vs. exchange time stamp for AAPL stock on August 11th, 2015.**

Table 1. Trades out of sequence for selected stocks on August 11th, 2015.

| Stock | Percent Out of Sequence | Total Trades | Listing Exchange |
|---|---|---|---|
| AAPL | 66.2% | 482,578 | NASDQ |
| BAC | 50.1% | 97,303 | NYSE |
| XOM | 43.0% | 86,834 | NYSE |
| GOOG | 56.3% | 69,085 | NASDQ |
| DNR | 43.3% | 51,185 | NYSE |
| IBM | 26.4% | 29,960 | NYSE |
| SHAK | 27.9% | 16,349 | NYSE |
| KLIC | 30.3% | 3,304 | NASDQ |
| GBX | 20.6% | 2,804 | NYSE |
| WBMD | 29.9% | 2,428 | NASDQ |
| EYES | 10.3% | 1,653 | NASDQ |
| BRKA | 0.3% | 304 | NYSE |
| OHGI | 3.8% | 286 | NASDQ |
| ACU | 0.0% | 1 | NYSEMKT |

## 5. CONCLUSIONS

Above, we uncover clear limitations to the accuracy of the SIP, namely its inability to preserve the correct ordering of trades. Rather than solely relying on a largely homogenous set of stocks (e.g., the components of the Dow 30), we instead opted for a heterogenous set of stocks which varied on market capitalization, listed exchange and trading volume. This heterogeneity yielded convincing evidence that the inaccuracy of the SIP relates to trade volume. Therefore, one can reasonably infer that the physical infrastructure underlying the SIP has finite limitations which are routinely exceeded for certain stocks.

While the impacts from such SIP inaccuracy could be extensive, we conclude by highlighting two significant implications. First, when the SIP fails to capture the actual order of trades, such inaccuracies will skew stock returns. While a simple measure, returns play a fundamental role in nearly all measures of risk; therefore, the inaccuracies of the SIP uncovered here could reveal extensive limitations for risk management. And finally, given the preponderance of algorithmic trading in today's U.S. equity markets, disordered trades and subsequently skewed returns could result in positive feedback driving markets away from efficient, price discovery.

While convincing, our findings on the accuracy of the SIP are not exhaustive, so we therefore encourage researchers to consider additional dimensions of analysis for future research. A longitudinal study might address the accuracy of the SIP over time. Another study might compare the accuracy of the SIP for component stocks of prevailing market indices (e.g., Dow 30, S&P 500 and Russell 3000). Lastly, we suggest that a direct extension of this study of the SIP accuracy of trades includes an analysis of updated quotes.

**Acknowledgments:** The authors gratefully acknowledge the following: *collaborative contributions* from Richard Bookstaber, Michael Foley, Christine Harvey, Eric Hunsader, Neil Johnson and Mark Paddrik; *helpful insights* from Anshul Anand, Chris Danforth, David Dewhurst, Peter Dodds, Jordan Feidler, Andre Frank, Bill Gibson, Frank Hatheway, John Ring, Chuck Schnitzlein, Colin Van Oort, Tom Wilk and attendees at the 2016 International Congress on Agent Computing. The authors' affiliation with The MITRE Corporation is provided for identification purposes only and is not intended to convey or imply MITRE's concurrence with, or support for, the positions, opinions or viewpoints expressed by the authors.

APPENDIX

Table A.1. Exchanges and Trade Reporting Facilities

| Exchange Family | Exchange Name | Abbreviations | Geographic Location |
|---|---|---|---|
| BATS | BATS | BATS / BZX | Secaucus, NJ |
| | BATS-Y | BATY / BYX | Secaucus, NJ |
| | Direct Edge A | EDGA | Secaucus, NJ |
| | Direct Edge X | EDGX | Secaucus, NJ |
| Chicago | Chicago Stock Exchange | CHX | Secaucus, NJ |
| NASDAQ | NASDAQ | NASD | Carteret, NJ |
| | NASDAQ-Boston | NQBS | Carteret, NJ |
| | NASDAQ-Philadelphia | NQPH | Carteret, NJ |
| NYSE | New York Stock Exchange | NYSE | Mahwah, NJ |
| | New York Stock Exchange - ARCA | ARCA | Mahwah, NJ |
| | New York Stock Exchange - Market | AMEX / NY-MKT | Mahwah, NJ |
| SIP | NYSE Trade Reporting Facility | NTRF | Mahwah, NJ |
| SIP | NASDAQ Trade Reporting Facility | QTRF | Carteret, NJ |